\documentclass[bibyear]{aa}

\usepackage{hyperref}
\usepackage{graphicx}
\usepackage{amsmath}
\usepackage{amssymb,bm}
\usepackage{tabularx}
\usepackage{longtable}
\usepackage{xcolor}
\usepackage{booktabs}
\usepackage[normalem]{ulem}
\usepackage{natbib}
\usepackage{float}
\usepackage{color}
\usepackage{enumitem}

\makeatletter
\renewcommand*\aa@pageof{, page \thepage{} of~\pageref*{LastPage}}
\hypersetup{colorlinks=true,allcolors=[rgb]{0,0,0.7}}

\begin{document}


\newcommand{\AH}[1]{{\textcolor{blue}{Amina:\,#1}}}
\definecolor{darkgreen}{rgb}{0,0.5,0}  
\newcommand{\TC}[1]{\textcolor{darkgreen}{#1}}
\newcommand{\changed}[1]{{\textcolor{black}{#1}}}
\newcommand{\smallspace}{\vspace{-.35cm}}
\newcommand*\diff{\mathop{}\!\mathrm{d}}


\newcommand{\eagle}{\textsc{eagle}}
\newcommand{\auriga}{\textsc{auriga}}
\newcommand{\Auriga}{\textsc{auriga}}
\newcommand{\agama}{\textsc{agama}}
\newcommand{\emosaics}{\textsc{emosaic}s}
\newcommand{\Au}{\textsc{Au}}
\newcommand{\MW}{\textsc{MW}}
\newcommand{\Gaia}{\textit{Gaia}}
\newcommand{\LCMD}{\ensuremath{\Lambda\mathrm{CDM}}}

\newcommand{\Disc}{\textit{Disc}}
\newcommand{\Halo}{\textit{Halo}}
\newcommand{\Ungrouped}{\textit{Ungrouped Halo}}
\newcommand{\GES}{\textit{GES}}
\newcommand{\HTD}{\textit{HTD}}
\newcommand{\HS}{\textit{HS}}
\newcommand{\Sequoia}{\textit{Sequoia}}
\newcommand{\LRLsixtyfour}{\textit{LRL64}}
\newcommand{\EDone}{\textit{ED1}}
\newcommand{\EDtwo}{\textit{ED2}}
\newcommand{\EDthree}{\textit{ED3}}
\newcommand{\EDfour}{\textit{ED4}}
\newcommand{\EDfive}{\textit{ED5}}
\newcommand{\EDsix}{\textit{ED6}}
\newcommand{\Thamnos}{\textit{Thamnos}}
\newcommand{\Typhon}{\textit{Typhon}}

\newcommand{\kpc}{\ensuremath{~\text{kpc}}}
\newcommand{\Mpc}{\ensuremath{~\text{Mpc}}}
\newcommand{\Gpc}{\ensuremath{~\text{Gpc}}}
\newcommand{\MpcCube}{\ensuremath{\,\text{Mpc}^{-3}}}
\newcommand{\MpcInv}{\ensuremath{\,\text{Mpc}^{-1}}}
\newcommand{\Msun}{\ensuremath{\,\text{M}_\odot}}
\newcommand{\kms}{\ensuremath{\,\text{km}\hspace{0.1em}\text{s}^{-1}}}
\newcommand{\uJ}{\ensuremath{\,\mathrm{kpc}\,\mathrm{km}/\mathrm{s}}}
\newcommand{\uE}{\ensuremath{\,\mathrm{km}^2/\mathrm{s}^2}}
\newcommand{\uEe}{\ensuremath{10^{5}\,\mathrm{km}^2/\mathrm{s}^2}}

\newcommand{\J}{\ensuremath{\bf{J}}}
\newcommand{\FobsJ}{\ensuremath{F_{\mathrm{obs}}\left(\mathbf{J}\right)}}
\newcommand{\FobsJcorrect}{\ensuremath{F_{\mathrm{obs}}^{*}\left(\mathbf{J}\right)}}
\newcommand{\FJ}{\ensuremath{F\left(\mathbf{J}\right)}}
\newcommand{\FJt}{\ensuremath{F\left(\boldsymbol{J},\boldsymbol{\theta}\right)}}
\newcommand{\FobsJt}{\ensuremath{F_{\mathrm{obs}}\left(\mathbf{J},\mathbf{\theta}\right)}}
\newcommand{\FobsJtcorrect}{\ensuremath{F_{\mathrm{obs}}^{*}\left(\mathbf{J},\mathbf{J}\right)}}
\newcommand{\SJ}{\ensuremath{S_{\mathrm{Obs}}\left(\mathbf{J}\right)}}

\newcommand{\Jr}{\ensuremath{J_{\textnormal{r}}}}
\newcommand{\JR}{\ensuremath{J_{\textnormal{R}}}}
\newcommand{\Jz}{\ensuremath{J_{\textnormal{z}}}}
\newcommand{\Jtotal}{\ensuremath{J_{\textnormal{total}}}}
\newcommand{\Lz}{\ensuremath{L_{\textnormal{z}}}}
\newcommand{\Met}{\ensuremath{\left[\mathrm{Fe}/\mathrm{H}\right]}}
\newcommand{\Emin}{\ensuremath{E_{\min}}}
\newcommand{\wJ}{\ensuremath{w\left(\boldsymbol{J}\right)}}
\newcommand{\tJ}{\ensuremath{\tau\left(\boldsymbol{J}\right)}}
\newcommand{\vtoomre}{\ensuremath{V_{\mathrm{Toomre}}}}
\newcommand{\vtan}{\ensuremath{V_{\mathrm{Tan}}}}
\newcommand{\vmax}{\ensuremath{V_{\max}}}
\newcommand{\vmed}{\ensuremath{V_{\mathrm{med}}}}
\newcommand{\pmin}{\ensuremath{p_{\min}}}

\newcommand{\FiJ}{\ensuremath{F_{i}\left(\mathbf{J}\right)}}
\newcommand{\FX}{\ensuremath{F\left(\mathbf{X}\right)}}
\newcommand{\X}{\ensuremath{\boldsymbol{X}}}

\newcommand{\Msol}{\ensuremath{M_{\odot}}}
\newcommand{\vlos}{\ensuremath{v_{\mathrm{los}}}}
\newcommand{\ud}{\ensuremath{\boldsymbol{u}^{5D}}}
\newcommand{\Mpeak}{\ensuremath{M_{\mathrm{peak}}}}
\newcommand{\Meight}{\ensuremath{ M_{\mathrm{peak}}>10^8M_{\odot} }}

\newcommand{\Act}{\ensuremath{ \left( \boldsymbol{J} , \boldsymbol{\theta} \right) }}
\newcommand{\Cart}{\ensuremath{ \left( \boldsymbol{x} , \boldsymbol{v} \right) }}
\newcommand{\sPDF}{\ensuremath{\sigma_{\mathrm{PDF}}}}

\newcommand{\NhsT}{\ensuremath{N^{\mathrm{HS}}_{\mathrm{Total}}}}
\newcommand{\NhsU}{\ensuremath{N^{\mathrm{HS}}_{\resizebox{0.1cm}{0.1cm}{{$\uparrow$}}}}}
\newcommand{\NhsD}{\ensuremath{N^{\mathrm{HS}}_{\resizebox{0.1cm}{0.1cm}{{$\downarrow$}}}}}

\newcommand{\alphaL}{\ensuremath{\alpha_{\boldsymbol{L}}}}


\title{Group Accretion in Milky Way-like Stellar Haloes}
\author{
Thomas M. Callingham
\and
Amina Helmi
}

\institute{
Kapteyn Astronomical Institute, University of Groningen, Landleven 12,
9747 AD Groningen, The Netherlands\\
\email{t.m.callingham@astro.rug.nl}
}


\date{}



\abstract{As galaxies form hierarchically,
  larger satellites may accrete alongside smaller companions in group infall events.
  This coordinated accretion is likely to have left signatures in the Milky Way's stellar halo at the present day.
 }
 {Our goal is to characterise the possible groups of companions that accompanied larger known accretion events of our Galaxy,
 and infer where their stellar material could be in physical and dynamical space at present day.
 }
 { We use the \Auriga{} simulation suite of Milky Way-like haloes to identify analogues to these large accretion events and their group infall companions, and we follow their evolution in time.
 }
 { We find that most of the material from larger accretion events is deposited on much more bound orbits than their companions. This implies a weak dynamical association between companions and debris, but it is strongest with the material lost first. As a result, the companions of the Milky Way's earliest building blocks are likely to contribute stars to the solar neighbourhood,
 whilst the companions of our last major merger are likely found in both the solar neighbourhood and the outer halo.
 More recently infallen groups of satellites, or those of a smaller mass,
 are more likely to retain dynamical coherence, for example, through clustering in the orientation of angular momentum.
 }
 {Group infall has likely shaped the Milky Way's stellar halo. Disentangling this will be challenging for the earliest accretion events, although overlap with their less-bound debris may be particularly telling.
 }


\maketitle



\section{Introduction}\label{Sec:Intro}
The current favoured cosmological model, $\Lambda$ Cold Dark Matter (\LCMD),
predicts that galaxies are assembled hierarchically; 
over time, larger galaxies grow in part by consuming smaller galaxies.
Evidence of this process can be found in the Milky Way's stellar halo,
which contains the material of destroyed \changed{dwarf galaxies} that fell into our Galaxy.
This process can be seen in action in the present day,
with several dwarf galaxies currently infalling, such as the Magellanic Clouds,
as well as Sagittarius, which is in the process of being fully accreted
\citep{ibata94DwarfSatelliteGalaxy}.

Older accretion events have long since phase phase-mixed;
their stars spread around the stellar halo in a diffuse physical distribution. 
However, their stars remain on similar orbits, following their progenitors \citep{johnston96FossilSignaturesAncient,helmi99BuildingStellarHalo}.
Hence, given a large enough sample of halo stars,
it is still possible to identify accretion groups based on their coherent dynamics.
Some groups, such as the Helmi Streams \citep{helmi99DebrisStreamsSolar}, have been long known, 
but the majority have been identified since \Gaia{} DR2 \citep{gaia-collaboration18GaiaDataRelease}.
Those found in \Gaia{} DR2 include our Galaxy's last major merger, Gaia-Enceladus-Sausage \citep{helmi18MergerThatLed,belokurov18CoformationDiscStellar};
an early building block, Kraken \citep[also referred to as Heracles,][]{kruijssen20KrakenRevealsItself,horta21EvidenceAPOGEEPresence};
and smaller accretion events such as Sequoia, Thamnos and RLR-64/Anteus
\citep{myeong18DiscoveryNewRetrograde,koppelman19MultipleRetrogradeSubstructures,ruiz-lara22SubstructureStellarHalo,oria22AntaeusRetrogradeGroup}.

In the most recent \Gaia{} dataset, DR3 \citep{vallenari23GaiaDataRelease}
smaller halo substructures  on the order of tens of stars have been identified,
such as the ED family of substructures \citep{dodd23GaiaDR3View},
alongside others \citep[e.g. Typhon,][]{tenachi22TyphonPolarStream}.
These smaller structures are possibly globular clusters, such as ED-2 \citep{balbinot2433BlackHole},
or the remnants of small ultra-faint galaxies.
Some of these smaller substructures, other streams
\citep[e.g.][]{malhan18STREAMFINDERNewAlgorithm, 
naidu20EvidenceH3Survey},
and ultra-faint dwarf galaxies in the outer halo \citep{simon19FaintestDwarfGalaxies}
have been suggested to be associated with larger accretion events
possibly as examples of group infall \citep{bonaca21OrbitalClusteringIdentifies,julio24SatelliteGroupInfall}.

Since hierarchical assembly  applies to galaxies of all masses, 
we expect the Milky Way (MW hereafter) to accrete dwarf galaxies that are themselves in the process of accreting
smaller galaxies and dark matter subhaloes.
These fall into the MW together;
a group infall event consisting of a primary satellite and its smaller companions \citep{li08InfallSubstructuresMilky}.
From simulation-based studies,
it has been suggested that around one-third of smaller subhaloes that infall onto MW-sized objects could have been
accreted as part of a group \citep{wetzel15SAtelLITEDWARFGALAXIES}.
The number of companions expected to accompany a primary satellite strongly depends on its mass
but with considerable scatter \citep[super-Poissonian; ][]{boylan-kolchin10TheresNoPlace}.
Whether these haloes are bright or dark depends on their mass and complex environmental factors,
such as the impact of reionisation \citep{moster13GalacticStarFormation,nadler19ModelingConnectionSubhalos,benitez-llambay20DetailedStructureOnset}.

Within our own Galaxy, the Magellanic Clouds provide a clear example of ongoing group infall;
the Large Magellanic Cloud (LMC) can be seen accreting with a close companion, the Small Magellanic Cloud (SMC), 
and possibly both ultra-faint dwarf galaxies 
\citep{sales11CluesMagellanicGalaxy,sales17IdentifyingTrueSatellites,kallivayalil18MissingSatellitesMagellanic}
and even brighter classical satellites \changed{\citep{donghia08SmallDwarfGalaxies,patel20OrbitalHistoriesMagellanic}}.
The infall of the Magellanic group has also been postulated as a possible explanation for the plane of satellites
\citep{lynden-bell95GhostlyStreamsFormation,pawlowski22EffectLargeMagellanic,taibi24PortraitVastPolar},
\changed{and group infall a possible mechanism for explaining observations of extragalactic satellite planes in general \citep{muller18WhirlingPlaneSatellite}.}
These possibilities can be studied by using simulations of MW-like galaxies to identify analogues to the LMC,
allowing predictions to be made for the expected population of bright companions and their dynamics
\changed{\citep{wang13SpatialDistributionGalactic,deason15SatellitesLMCmassDwarfs,dooley17PredictedLuminousSatellite,santos-santos21MagellanicSatellitesLCDM}}.

Almost by definition, infall groups enter the halo around the same time \citep{dsouza21InfallDwarfSatellite}
and with a similar orientation of angular momentum \citep{li08InfallSubstructuresMilky}. 
However, whilst we know that satellites infall together, they may not stay together in phase-space.
Infall groups such as the Magellanic system are loosely bound
and typically lose dynamical coherence, spreading in both position and velocity space 
over $4$ to $8$ Gyrs \citep{deason15SatellitesLMCmassDwarfs,taibi24PortraitVastPolar}.

Several works have studied the initial group infall of satellites \citep{dsouza21InfallDwarfSatellite}
and their following destruction \citep{trelles22ConcurrentInfallSatellites}.
Less attention has been paid to the relationship between the stellar material of the group at the present day, 
to whether the stellar material of the companions occupies the same radial ranges as their primary satellites,
or if their material is dynamically similar enough to be related to their primary hosts in orbital space. 
Given our current knowledge of the more significant accretion events in the MW's assembly history,
what companions should we expect to have accompanied them?
Can their material still be associated with that of their original primary host, or have they forgotten their companionship?

In this paper, we use the \Auriga{} simulations to study group infall
and explore possible implications for the stellar halo of our Galaxy.
We identify analogues to the MW's key accretion events and classify their group infall companions.
We examine these populations of companions, both bright and dark,
characterising their (dynamical) properties. 
We follow these groups through their accretion and orbital evolution to survival or destruction.
Finally, we consider the distribution of their material in physical and dynamical space in the present day
and how the companion's material relates to that of the primary satellite.
We are principally interested in older group infall events,
where the primary satellite has long since been destroyed and accreted.

The structure of the paper is as follows.
In Sec.~\ref{Sec:AuData}, we introduce the \Auriga{} simulations and our definition for group infall.
In Sec.~\ref{Sec:MW_data}, we define selections of analogues to MW accretion events
and examine the expected populations of their companions.  
In Sec.~\ref{Sec:GI_ER}, we look at where the material of accreted groups reside in physical space and energy at the present day and 
in Sec.~\ref{Sec:GI_IoM}, we explore their orbital coherence.
In Sec.~\ref{Sec:MW_Discuss}, we discuss the implications for the MW and present our conclusions in Sec.~\ref{Sec:Conclusion}.

\section{Group Infall in Auriga}\label{Sec:AuData}
\begin{figure}
 \includegraphics[width=\columnwidth]{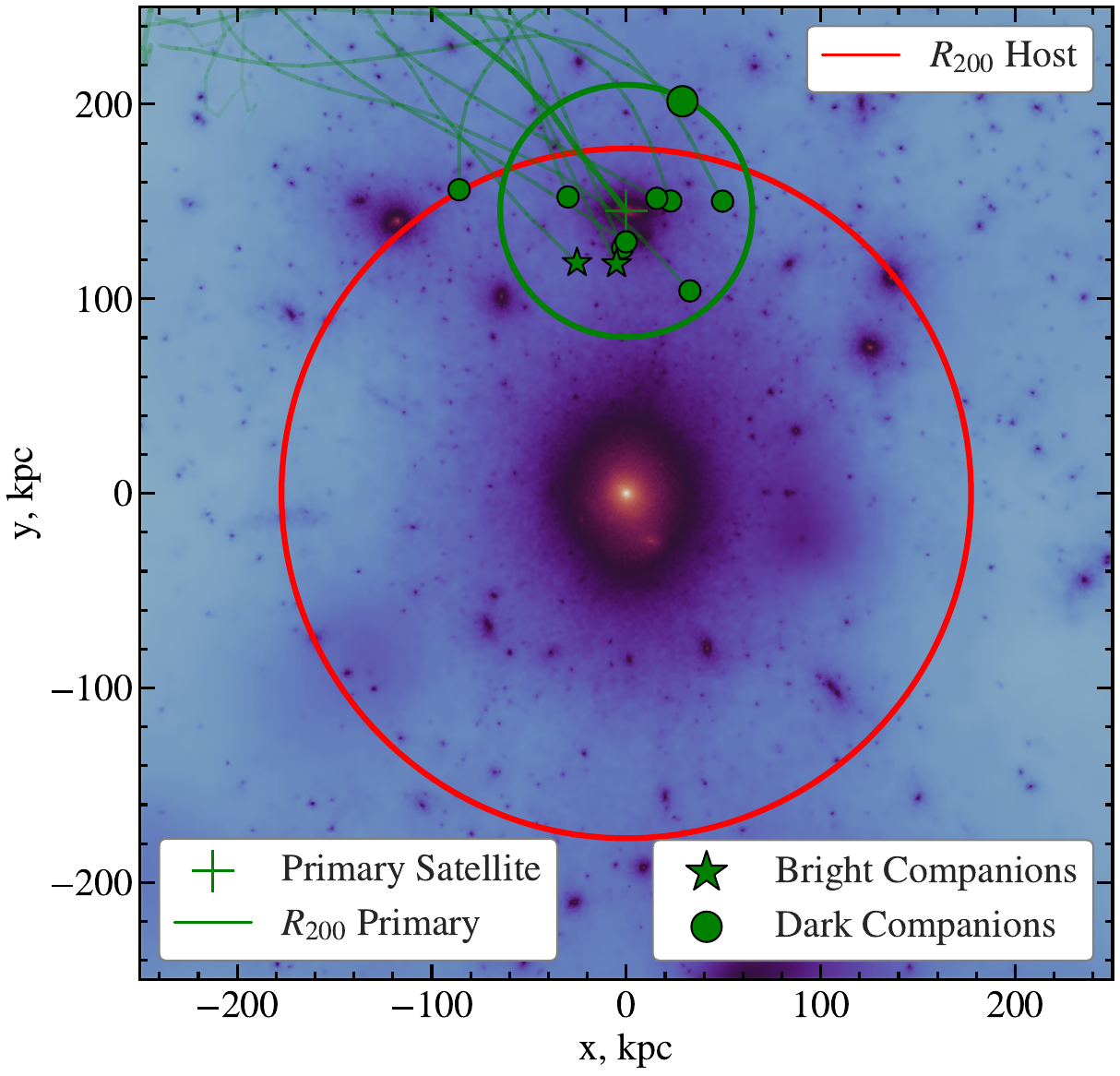}
\vspace{-12pt}
 \caption{An example of group infall in halo 27 of the Level-4 \Auriga{} simulations 
 \citep[made with Pynbody;][]{pontzen13PynbodyNBodySPH}.
 We define group infall as companion satellites  within the $R_{200}$ of the primary satellites
 when the primary satellite first touches the central MW-like host galaxy (see text for details).
 This group is plotted at the snapshot immediately after primary infall.
\vspace{-10pt}
 }\label{fig:GI_Pretty}
\end{figure}

\begin{figure*}
 \includegraphics[width=\textwidth]{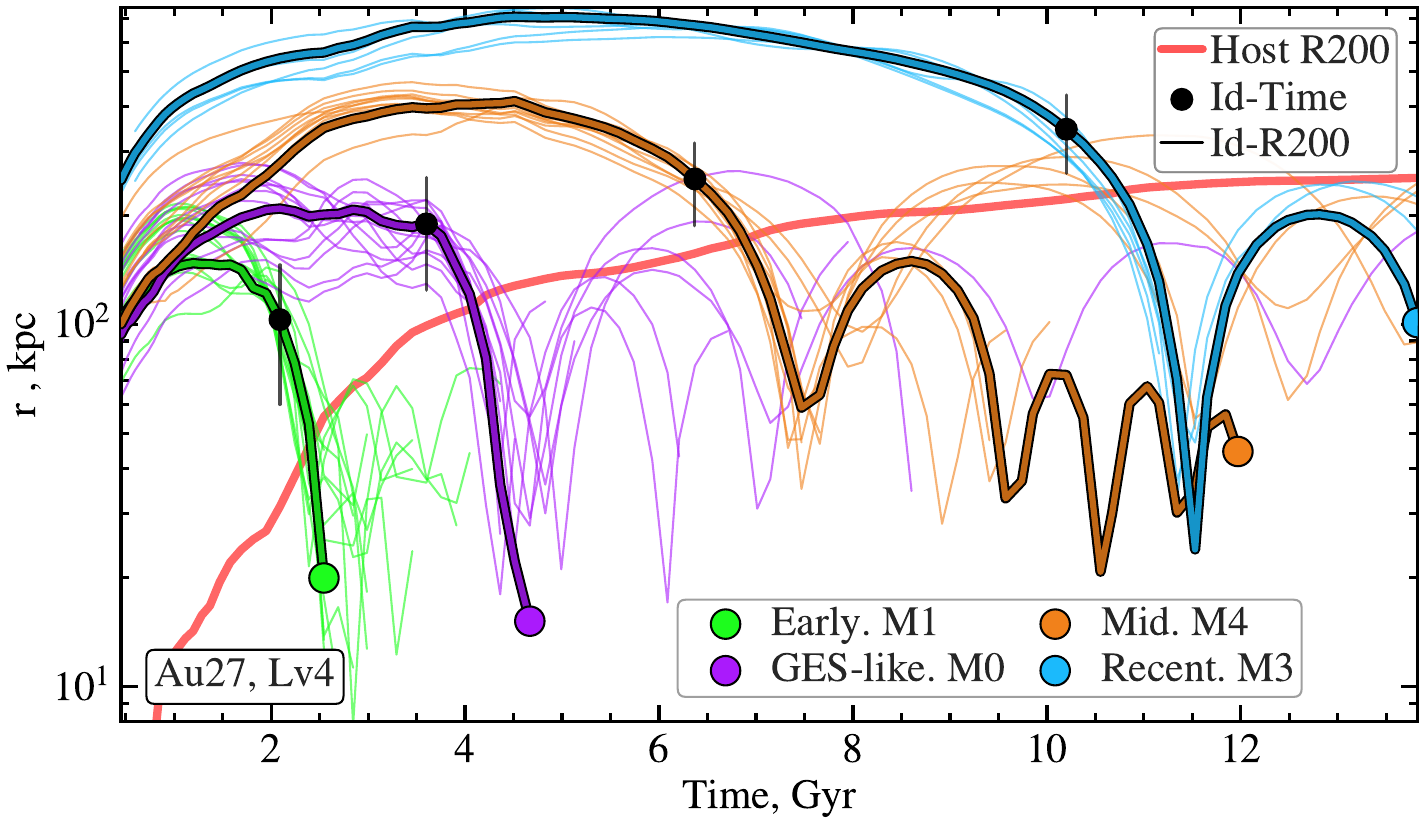}
\vspace{-10pt}
 \caption{Group infall of different accretion events, 
 depicted as the radial distance of the satellites against time.
 The $R_{200}$ of the central MW-like galaxy is marked in thick red.
 The different group infall events are represented with different colours,
 with the orbits of the primary satellites in thicker solid lines
 and their smaller companions in thinner transparent lines.
 The identification time of the group and the $R_{200}$ of the primary satellite is marked in black.
 Smaller subhaloes within the $R_{200}$ of the primary satellite at this time are considered companions.
 These accretion events are selected as analogues to events in the MW, as described in  Sec.~\ref{Sec:MW_data}.
\vspace{-10pt}
 }\label{fig:GI_TimeR}
\end{figure*}

We use the MW-like high-resolution cosmological zoom-in simulations of the \auriga{} project \citep{grand17AurigaProjectProperties},
a suite of 30 haloes with masses between $1-2\times 10^{12} M_{\odot}$.
These were selected from the $100^{3} \, \textnormal{Mpc}^{3}$ periodic cube of the \eagle{} project,
a $\Lambda$CDM cosmological hydrodynamical simulation \citep{crain15EAGLESimulationsGalaxy,schaye15EAGLEProjectSimulating}
adopting Planck1 cosmological parameters \citep{planckcollaboration14Planck2013Results}.
These selected haloes were individually resimulated as a zoom-in simulation
using the N-body and moving mesh magnetohydrodynamic \textsc{arepo} code \citep{springel11MovingmeshHydrodynamicsAREPO}.
We here use the level 4 resolution sample,
with a DM particle mass of $4.1\times 10^{5} M_{\odot}$ and
an initial gas resolution element of mass $5.1\times 10^{4} M_{\odot}$.
We label the haloes Au$1$ to Au$30$.

The DM haloes are identified using the Friends-of-Friends algorithm  \citep[FoF,][]{davis85EvolutionLargescaleStructure}.
The subhaloes within the FoF groups are found using Subfind \citep{springel05CosmologicalSimulationCode}.
In this work, we recenter the simulations using a shrinking spheres algorithm
and rotate the simulations to the orientation of the disc at the present day.
Infall is defined as the time that a subhalo first crosses the virial radius ($R_{200}$) of the main host galaxy,
estimated by interpolating the radial position of the satellite between snapshots.
We define accreted subhaloes as those which have fallen in (crossed $R_{200}$ of the main host)
and subsequently either have been destroyed onto the main halo or survive at the present day within the $R_{200}$ of the main halo.
Subhaloes are considered destroyed where they are no longer found in the merger tree.

Like any galaxy simulation, the \Auriga{} project has a finite resolution,
which limits the size of subhaloes that will be confidently resolved.
\citet{grand21DeterminingFullSatellite} estimated this limit to be around 100 particles, 
corresponding to $\sim4\times10^7\Msol$ for level 4. 
Furthermore, the \Auriga{} simulations are known to underestimate the number of bright satellites in the low-mass,
ultra-faint regime \citep{grand21DeterminingFullSatellite}.
We thus consider the number of bright satellites and bright companions predicted by \Auriga{} to be a lower limit.
An upper limit to the number of bright satellites can be obtained by assuming all subhaloes of a peak total mass ($\Mpeak$) greater than $10^{8}M_{\odot}$ are bright 
\citep[following][]{deason15SatellitesLMCmassDwarfs,garrison-kimmel14ELVISExploringLocal}.
Nonetheless, for completeness, we follow all companion subhaloes. 
To track the likely distribution of the stars/particles from the smaller subhaloes,
we consider the most bound particle identified at infall time 
\citep[as discussed in][]{grand21DeterminingFullSatellite,santos-santos25UnabridgedSatelliteLuminosity}.

\subsection{Definining Group Infall}

In the literature, there are several definitions of group infall.
\citet{li08InfallSubstructuresMilky} used a definition based on Angular Momentum at the moment of infall,
whilst
\citet{deason15SatellitesLMCmassDwarfs} used a definition based on the group labels of the \textsc{Rockstar} halofinder algorithm
\citep{behroozi13ROCKSTARPhasespaceTemporal}.
\citet{dsouza21InfallDwarfSatellite} used a broader selection based on infall time
but then also considered and compared definitions of \citet{deason15SatellitesLMCmassDwarfs} and dynamical arguments.

In this work, we follow a similar definition to \citet{santos-santos21MagellanicSatellitesLCDM},
originally used for LMC-analogues but adapted to become more general.
The process is as follows:
\begin{enumerate}
 \item The identification time is set by the snapshot preceding a primary satellite first touching the host MW halo:\\ 
 $r_{\mathrm{sat}}< R^{\mathrm{Host}}_{200} + R^{\mathrm{Sat}}_{200}$,
  where $r_{\mathrm{sat}}$ is the distance to the primary satellite from the centre of the host MW halo.
 \item Subhaloes within this primary satellite's $R^{\mathrm{Sat}}_{200}$ at this identification time,
 \changed{and in the preceding snapshot,}
 are considered companions of the primary if they are not already companions of a larger primary satellite.
\end{enumerate}
By deliberate choice, this definition is likely to be conservative compared to other definitions made in other works.
Accounting for resolution effects,
we find that \changed{$\sim20\%$} of subhaloes with mass less than $5\times10^{9}\Msol$ have accreted as part of a group in Auriga level-4,
consistent with estimates from other works \citep{li08InfallSubstructuresMilky,wetzel15SAtelLITEDWARFGALAXIES}.

Other subhaloes not classified as companions, which we refer to as isolated satellites,
could be bound to, or gravitationally focused by, the primary \citep{dsouza21InfallDwarfSatellite}.
However, to maximise the possibility of dynamical coherence in the present day,
we are interested in only the closest companions, motivating our selection.

\subsection{An Example of Group Infall}

\begin{figure}
 \includegraphics[width=\columnwidth]{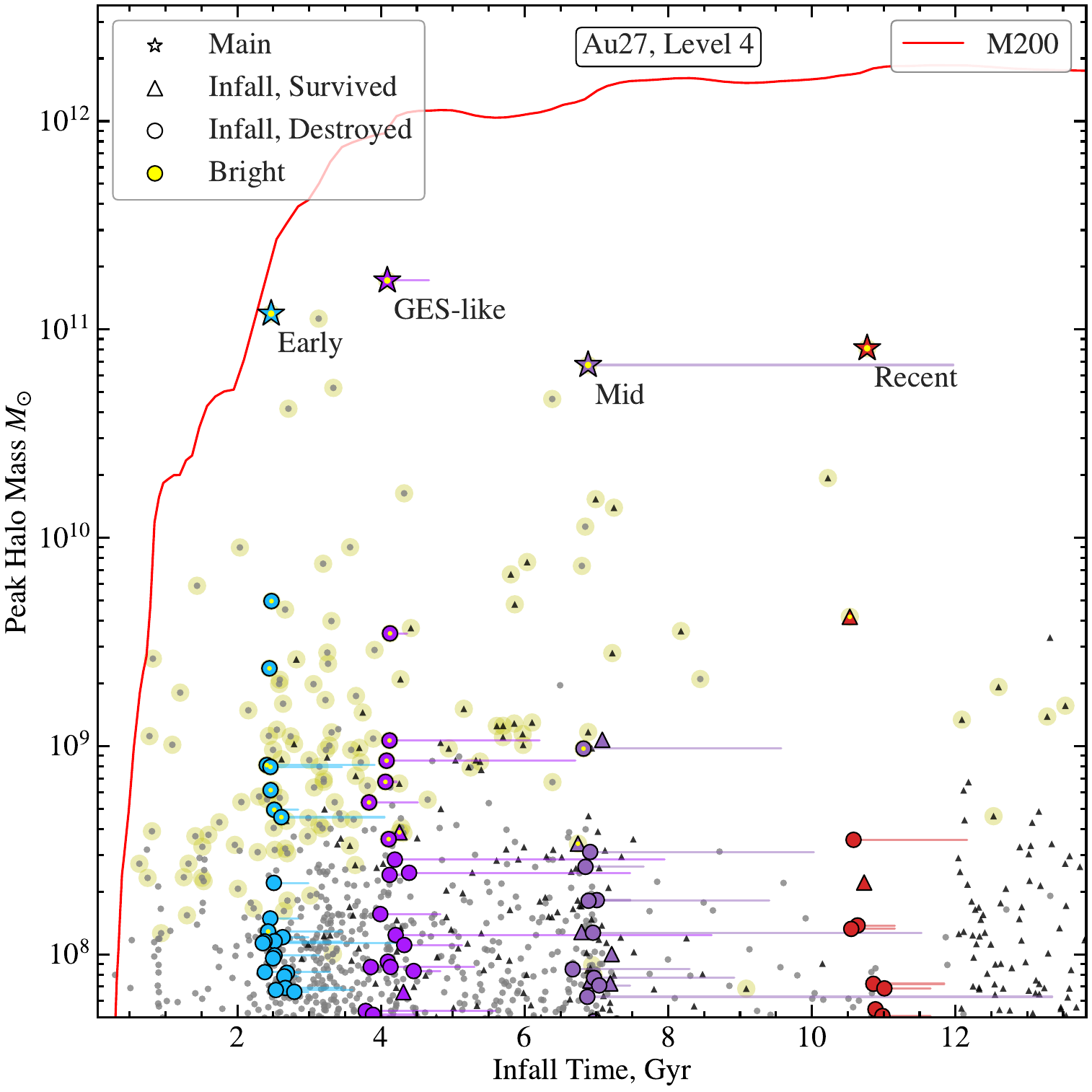}
\vspace{-12pt}
 \caption{Accretion history of \Auriga{} Halo 27 (resolution level 4),
 in the space of infall time and peak halo mass, where points represent accreted subhaloes.
 Subhaloes that survive to the present day are marked with a triangle,
 whilst circles denote those destroyed by the present day.
 Such subhaloes additionally have a line indicating how long they survived after infall.
 An additional yellow colouring represents bright satellites (subhaloes that have contained stars).
 Different infall groups are presented in different colours,
 selected by similarity to MW accretion events (see Sec.~\ref{Sec:MW_data} for details).\\
 \vspace{-10pt}
 }\label{fig:GI_MassTime}
\end{figure}

An example of group infall in \Auriga{} can be seen in Fig.~\ref{fig:GI_Pretty},
where the companions are associated with the primary satellite.
This is made clearer by considering the orbits of groups as radial distance against time in Fig.~\ref{fig:GI_TimeR}.
It can already be seen across the different example group infall events that 
companions can outlive and suffer a different fate to their respective primary satellite.

The accretion history of a single halo can be seen in Fig.~\ref{fig:GI_MassTime},
where the examples of group infall are highlighted by colour.
As expected, most companions infall at a time similar to the primary.
It is also possible to see that the distribution of subhaloes is clumped in time around the largest accretion events,
as observed by \citet{dsouza21InfallDwarfSatellite}.
The selected accretion events are motivated in the following section.

\section{Analogues to MW Accretion Events}\label{Sec:MW_data}

\begin{figure}
 \includegraphics[width=\columnwidth]{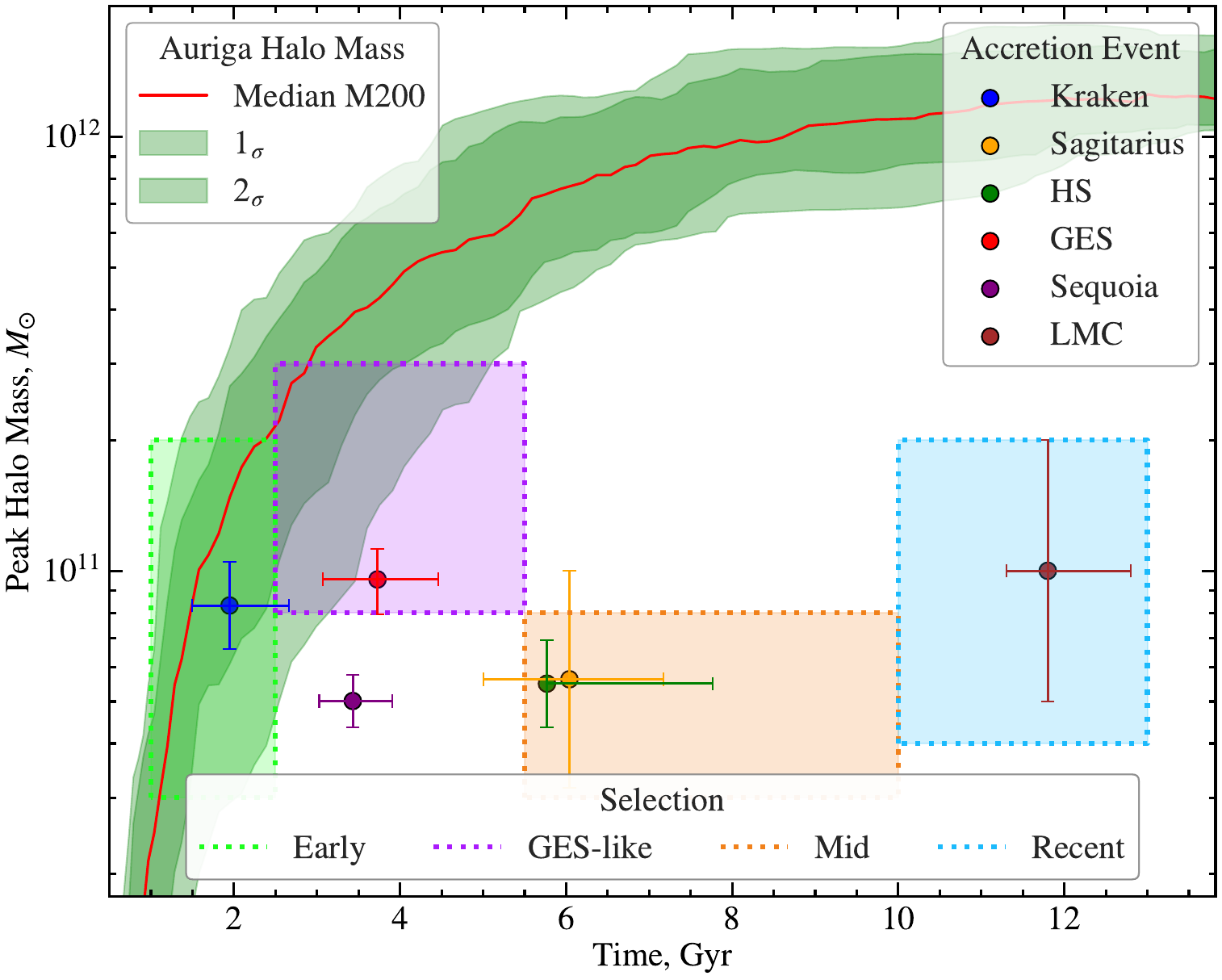}
\vspace{-12pt}
 \caption{
 Accretion history of an \changed{``average''} Milky Way, in the space of lookback time and peak halo mass.
 We have marked some important events in the accretion history of the MW known to date,
 with scatter points and uncertainties based on values from the literature (see text for details).
These motivate regions of interest in this space,
chosen to capture the characteristics of some of the key accretion events,
with 4 selections following Table~\ref{tab:MWSelection}.
 The red line is the median $M_{200}$ mass of the main halo across the level 4 \Auriga{} simulations against time,
 with green shaded regions indicating the $1\sigma$ and $2\sigma$ spreads (defined as the $18-84\%$ and $5-95\%$ ranges).
 }\label{fig:MW_TMass}
\vspace{-10pt}
\end{figure}

To make effective comparisons between the simulations and the MW,
it is first helpful to identify accretion events in the simulations that are similar to those that our own Galaxy has experienced.
The likely main accretion events in the MW's assembly history are:
Kraken \citep[although see][]{horta21EvidenceAPOGEEPresence}, GES, Sagittarius, and the LMC.
In the literature, there are various estimates of these accretion events infall time and total mass
with large uncertainties \citep[such as][]{kruijssen20KrakenRevealsItself,callingham22ChemodynamicalGroupsGalactic}.
These can be used to define four corresponding epochs: ``Early'', ``Last Major Merger'', ``Middle''  and ``Recent'',
indicated in Fig.~\ref{fig:MW_TMass} as box selections in infall time and peak halo mass.

\subsection{Definitions}
Here, we describe our selections in infall time and peak halo mass.
Additionally, we require that the subhaloes are primary satellites and not already companions to another halo.
A summary of the criteria can be found in Tab.~\ref{tab:MWSelection},
which also includes the resulting statistics of the selections,
such as the total number of analogues, the average number of bright companions,
and the fraction of both that survive to the present day.
An example of these selections in an \Auriga{} halo and their corresponding infall groups can be seen in the space of time and peak mass in Fig.~\ref{fig:GI_MassTime},
and orbits against time in Fig.~\ref{fig:GI_TimeR}.
\changed{Further examples of the orbits of the selected analogues can be seen in Appendix~\ref{App:ExampleSelections}.}

\begin{table*}[]
 \centering
 \caption{Our selections of MW-like accretion events, defined on infall time  and peak halo mass of the main satellite
  and visualised in  Fig.~\ref{fig:MW_TMass} (see text for details).
  The average number of bright companions column gives the median and $16-84\%$ ranges
  of the number of companions that have hosted a star particle in \auriga.
  Alternatively, the average number of companions with a peak mass greater than $10^{8}M_{\odot}$
  can be thought of an approximate upper bound for subhaloes that could plausibly host stars. 
  The last column gives the average number that survives to present day
   }
 \label{tab:MWSelection}
 \begin{tabular}{l|c|c|c|c|c|c|c|c} 
     & Infall Time & Halo Mass      & MW & 
     &  \multicolumn{2}{c|}{Av No. Companion} & \multicolumn{2}{c}{Av No. Comp Survive} \\
  Selection & (Gyr)  & ($\times{10}^{10}\,M_{\odot}$) & Analogues &
  Count &  Bright & \Meight & Bright & \Meight \\[2pt]
  \hline

  Early & $\left[1-2.5\right]$ & $\left[3-20\right]$ & Kraken &
   10 &  $4.5^{+4.5}_{-3.5}$ & $6.5^{+7.8}_{-4.5}$ & $0.0^{+0.0}_{-0.0}$ & $0.0^{+0.0}_{-0.0}$ \\[2pt]

  Major & $\left[2.5-5.5\right]$ & $\left[8-30\right]$ & GES &
  8 & $4.5^{+3.5}_{-2.4}$ & $13.5^{+11.9}_{-5.8}$ & $0.5^{+1.0}_{-0.0}$ & $1.0^{+1.9}_{-0.9}$ \\[2pt]

  Mid & $\left[5.5-10\right]$ & $\left[3-8\right]$ & Sagittarius &
   18 & $1.0^{+1.0}_{-0.0}$ & $3.0^{+4.3}_{-2.0}$ & $0.0^{+1.0}_{-0.0}$ & $1.^{+3.0}_{-0.}$ \\[2pt]

  Recent & $\left[10-13\right]$ & $\left[4-20\right]$ & LMC &
  5 & $1.0^{+1.4}_{-0.}$ & $7.0^{+0.4}_{-2.7}$ & $1.0^{+2.4}_{-0}$ & $5.0^{+1.4}_{-3.0}$ \\
\end{tabular}
\end{table*}

\subsubsection*{\underline{Early Building Blocks}}
Although debated, the earliest known significant accretion event in our Galaxy is likely Kraken
\citep{massari19OriginSystemGlobular,kruijssen20KrakenRevealsItself},
an ancient building block of the MW.
We define a selection between $1-2.5$ Gyrs and $3-20\times10^{10}\Msol$.
If multiple groups are selected, we take the largest selected accretion event.
We find 10 analogues in \Auriga{} level 4.

\subsubsection*{\underline{Last Major Merger}}
The last major merger of the MW is Gaia-Enceladus-Sausage (GES),
which likely accreted over 10 Gyrs ago with an approximate halo mass of $1\times10^{11}M_{\odot}$ \citep{helmi18MergerThatLed,deason19TotalStellarHalo,callingham22ChemodynamicalGroupsGalactic}.
We define a selection between $2.5-5.5$ Gyrs and $8-30\times10^{10}M_{\odot}$
and additionally require this to be the largest destroyed accretion event.
We find 8 analogues in \Auriga{} level 4.

\subsubsection*{\underline{Middle Events}}
These are satellites accreted after the last major merger, but over $4\,\mathrm{Gyrs}$ ago,
such as Sagittarius or the Helmi Streams in the MW \citep[e.g.][respectively]{law10SagittariusDwarfGalaxy,woudenberg24FirstMeasurementTriaxiality}.
We define a selection between $5.5-10$ Gyrs and $3-8\times10^{10}M_{\odot}$.
Additionally, we require these accretion events to infall after the main galaxy's last major merger.

We find 18 analogues in \Auriga{} level 4, with a considerable variety of satellites
with different orbits, survival times, and companion populations.
Half of these primary satellites are destroyed,
whilst the rest survive to the present day on more circular, slowly inspiring orbits.

\subsubsection*{\underline{Recent Infall Events}}
These are LMC-like analogues in the MW,
surviving objects that have only just infallen.
We define a selection between $10-13$ Gyrs and $4-20\times10^{10}M_{\odot}$,
and require that the subhalo survives to the present day.
We find 5 analogues in \Auriga{} level 4, all of which have completed one pericentric passage.

\subsection{Companion Populations}

\begin{figure}
 \includegraphics[width=\columnwidth]{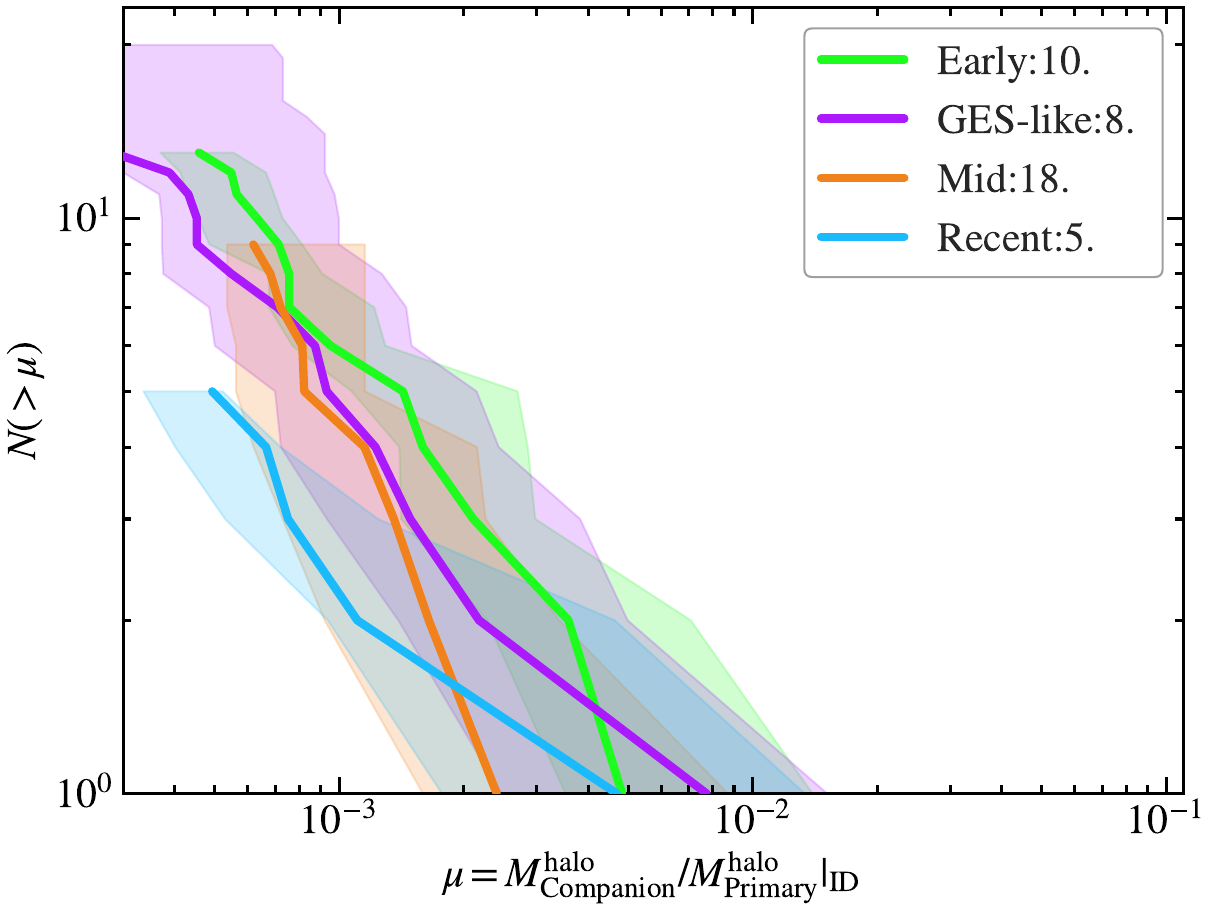}
\vspace{-12pt}
 \caption{Distributions of the companions' halo mass as a fraction of the primary satellite's at the time the group is identified;
 when the primary satellite first touches the main halo (see text for details).
 The solid lines correspond to the median across our samples, with the shaded region giving the $1\sigma$,
 defined by the $16\%-84\%$ of the distribution.
 Different colours correspond to our different analogue selections.
\vspace{-10pt}
 }\label{fig:MassDist}
\end{figure}

With our analogue accretion events identified, we now turn to study their population of companions of group infall.
The companion-primary mass distributions for the different selections can be seen in Fig.~\ref{fig:MassDist}.
\changed{
We find no statistical evidence for systematic differences in mass distributions between the selections,
with relatively high scatter in the individual distributions.
}
\changed{What is arguably more relevant to the history of the Milky Way is the size of the largest companions within the group.
As demonstrated by the SMC  in the Magellanic group 
\citep[$M_{*}\sim5\times10^{8}\Msol$][]{mcconnachie12OBSERVEDPROPERTIESDWARF}
massive primary satellites can be accompanied by a significant bright satellite.}
In peak stellar mass,
$75\%$ of Early accretion events have a companion that was larger than $1/100$,
$25\%$ of GES-like events,
$16\%$ of Mid events,
and a third of the recent LMC-like accretion events\footnote{\changed{In Appendix \ref{app:B}, we do a resolution study and show that the largest companion of a primary satellite with halo mass $10^{10} $ M$_\odot$ has converged in our simulations.}}. 
\changed{To summarise the analogues companion populations:}
\begin{itemize}
\item \textbf{Early:}
These early accretion events typically have between \changed{5 and 7}  bright companions,
using our lower estimate of \Auriga{}'s count of bright satellites and upper estimate of the number of satellites $\Mpeak>10^{8}\Msol$.
None of these companions survives to the present day.
\item \textbf{GES:}
These major accretion events typically have \changed{5 to 14} bright companions,
with a 1 to 3 surviving until the present day for our resolution.
\item \textbf{Mid:}
These objects typically bring in \changed{1 to 3} bright companions,
which can survive alone or be closely associated with the surviving primary satellite.
\item \textbf{Recent:}
These objects typically bring in a single bright companion in our \Auriga{} sample.
However, counting satellites with $\Mpeak>10^{8}M_{\odot}$, this increases to \changed{5} on average,
bringing the expected population more in line with the results of other studies \citep{deason15SatellitesLMCmassDwarfs}.
\end{itemize}

\section{Location of Companions within the Galaxy}\label{Sec:GI_ER}
\begin{figure}
 \includegraphics[width=\columnwidth]{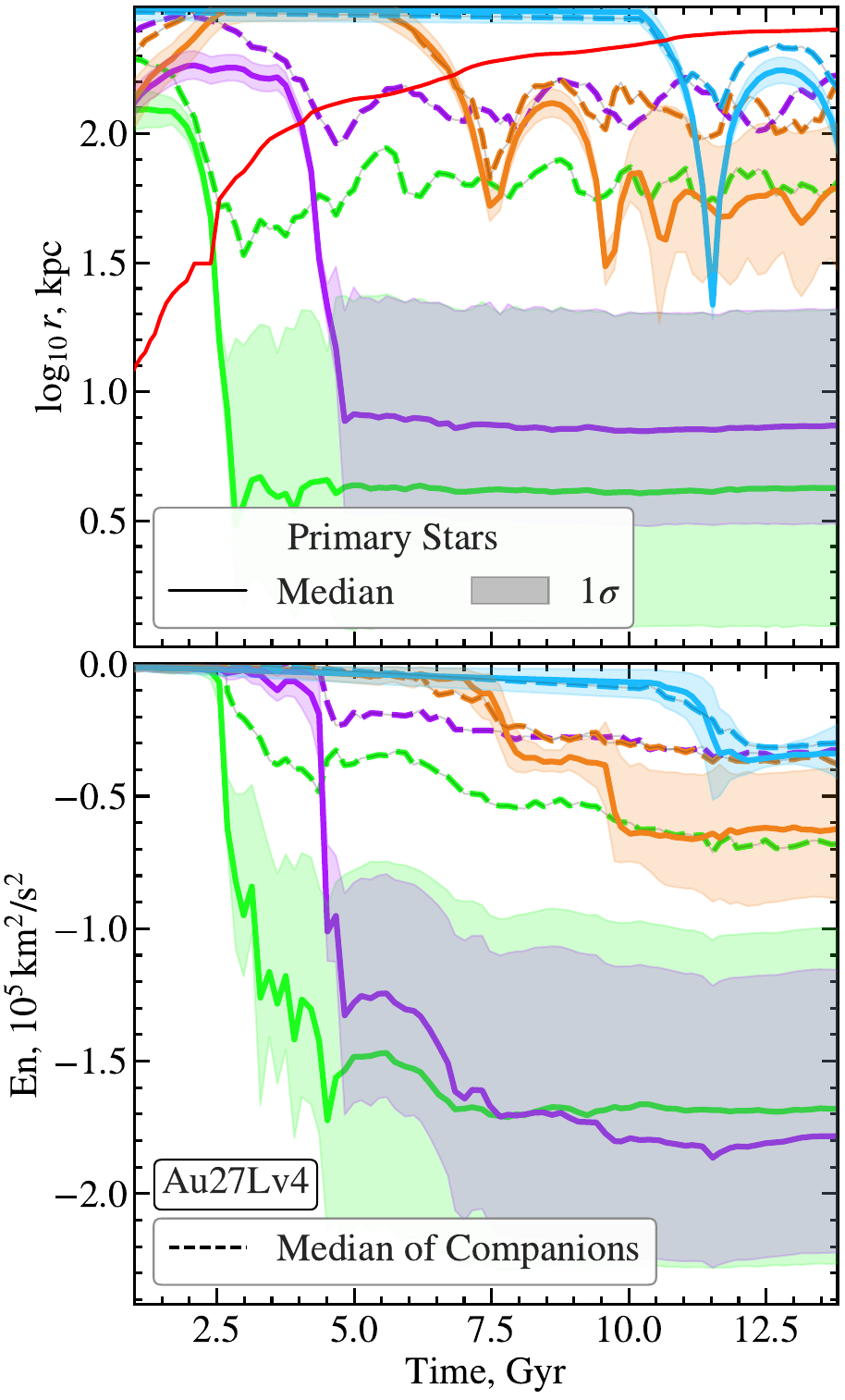}
\vspace{-12pt}
 \caption{Evolution in radial distance and energy (top and bottom panels, respectively)
 for examples of our selected analogues (different colours) over time
 \changed{ in \Auriga{} halo 27}.
 The solid lines correspond to the median of the  stellar material of the accretion event, with the shaded region giving the $1\sigma$,
 defined by $16\%-84\%$ of the distribution.
 The dashed lines represent the median of the subhalo companions of group infall, tracked by following the most bound particle at infall.
\vspace{-10pt}
 }\label{fig:Accretion}
\end{figure}

In the present day, the position of an accreted subhaloes material within the stellar halo
depends on the history of its progenitor,
the initial accretion and the following stripping and destruction.
In turn, this has a complex dependence on the progenitor's infall time, mass, and orbit \citep{amorisco17ContributionsAccretedStellar}.
However, there are a few general trends.

In  Fig.~\ref{fig:Accretion}, 
we follow our selected examples of group infall \changed{for Au27} in radius and energy over time,
considering both the primary satellite's stellar distribution and its companions' median. 
To track the central distributions of the companions after the subhaloes' destruction (and disappearance from the merger trees),
we follow the most bound particle at infall.
To compare the radius within different-sized \Auriga{} haloes,
we scale radial values to the MW using a mass-based scaling factor:
$\sqrt[3]{\frac{M^{\mathrm{MW}_{200}}}{M^{\mathrm{Au}_{200}}}}$,
using $M^{\mathrm{MW}_{200}}=1.17\times10^{12}\Msol$ \citep{callingham19MassMilkyWay}.

The material of earlier accretion events can, with scatter, be found at lower (higher binding) energy,
corresponding to the galaxy's inner regions,
\changed{whereas the material of later accretion events typically orbit at higher energy and radius}.
As a massive satellite orbits within the DM halo of the host system, it suffers dynamical friction, 
losing orbital energy and sinking further in. 
The least-bound material, the outer layers of DM, are stripped first.
The central stellar distribution survives longer and are often stripped through successive pericentres 
as the subhalo sinks further into the host,
depositing its innermost material closest to the galaxy's centre and at high binding energy.

\begin{figure}
\includegraphics[width=\columnwidth]{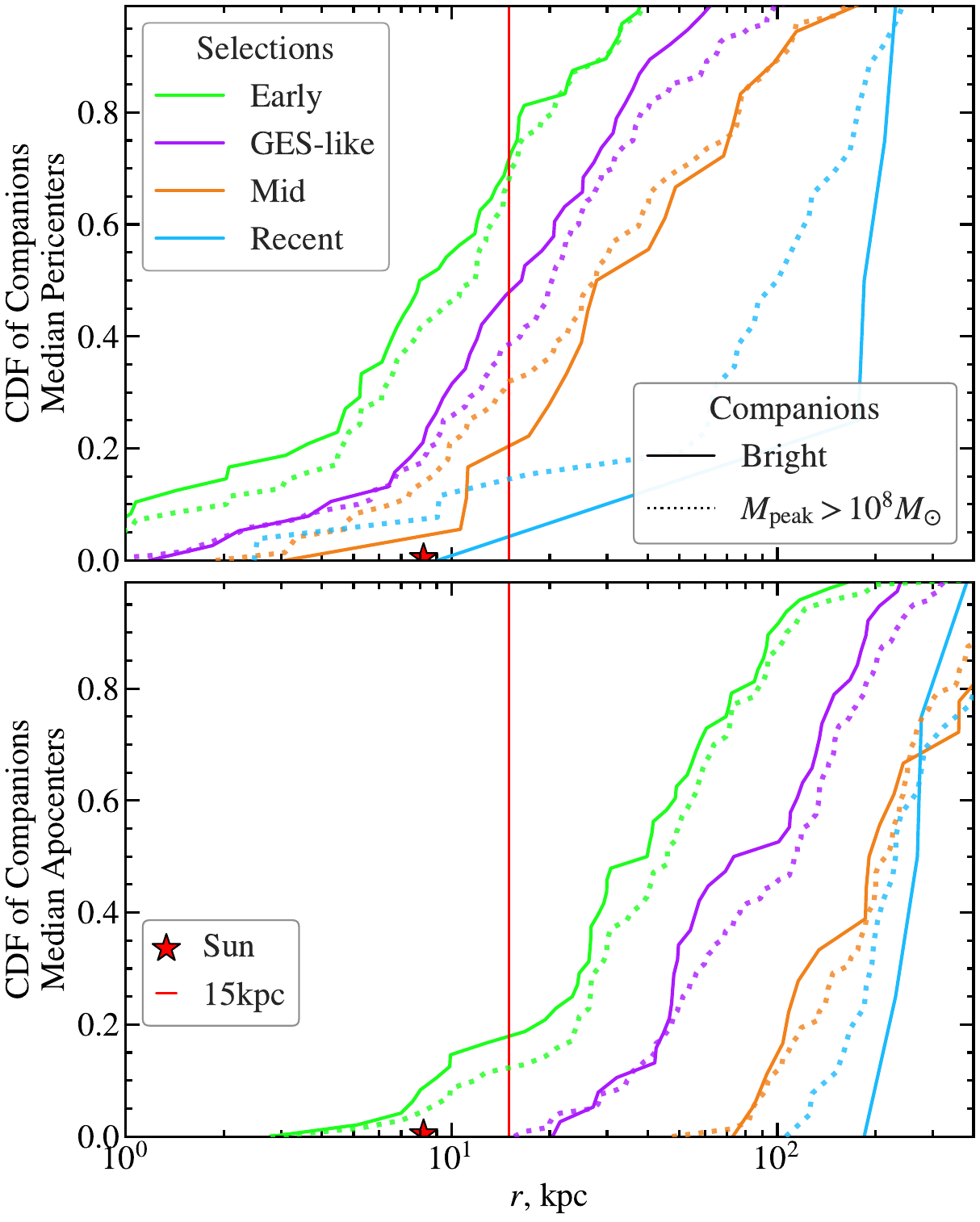}
\vspace{-12pt}
 \caption{
  CDF of the pericentres (top) and apocentres (bottom) of the material of the companions of group infall events,
  represented by the most bound particle to the subhalo at infall.
 The groups are classified by our analogue selection (see text for details) presented in different colours.
 Two subselections of companions are plotted in different linestyles;
 those that are bright (contain a star particle) in a solid line,
 and those with a peak total mass greater than $10^{8}M_{\odot}$ that could be bright in dotted.
 To compare the radius within different-sized \Auriga{} haloes,
 we scale radial values to the equivalent MW value (see text for details).
The vertical red line shows $15$~kpc, a generous limit of what can be considered the inner regions and the local stellar neighbourhood.
 }\label{fig:cdf_peri}
\vspace{-10pt}
\end{figure}

Typically, the companion satellites are in the outer regions of the primary satellite before the initial group infall. 
During accretion, they are stripped along with the outer layers of the subhalo
and so can typically be found at a larger radius and higher (low binding) energy.
Whilst the groups fall in together, they do not stay together spatially long.
Hence, for larger primary satellites that experience dynamical friction,
the companions are ``left behind'', with a smaller overlap with the bulk of the stellar material.


\changed{In our own Galaxy,
our ability to observe accreted debris is limited by our relative distance to the stellar material,
which depends on the radial range within the material orbits.
This range can be approximately described by the pericenters and apocenters of the most bound particles of the companions,
as shown for our selected analogues in Fig.~\ref{fig:cdf_peri}.
To pass through the solar neighbourhood,
a companion's pericenter (see top panel) must lie within the solar radius.
However, most of the phased mixed accreted material will be found near the apocenter of their orbits (see bottom panel).
The distributions of bright companions (solid) and companions with a peak mass greater than $10^{8}M_{\odot}$ (dotted) are similar,
suggesting that there is no strong dependence on mass, given that the companions are sufficiently small.}

\changed{
Nearly $90\%$ of the pericenters of earliest companions lie within 15~kpc,
suggesting that the stellar debris of bright companions of Kraken-like events can likely be found in the solar neighbourhood.
In contrast,
roughly half of the companions of GES-like accretion events have pericentres beyond this region,
placing most of the companions in the outer stellar halo.
Based on our estimates of the likely bright population of GES-like events (see Table~\ref{tab:MWSelection}),
we predict that the debris of $2-7$ bright satellites will be found beyond $15$~kpc.
For the companions of the Mid-class satellites,
the radial ranges vary significantly depending on the specific accretion event,
with a median pericentre $\sim$30~kpc.
On the other hand, the satellites of the most recent LMC-like accretion events are probably still in the distant halo,
with most having completed a single pericentre.
}

\section{Orbital Coherence of Infall Groups}\label{Sec:GI_IoM}
\begin{figure}
 \includegraphics[width=\columnwidth]{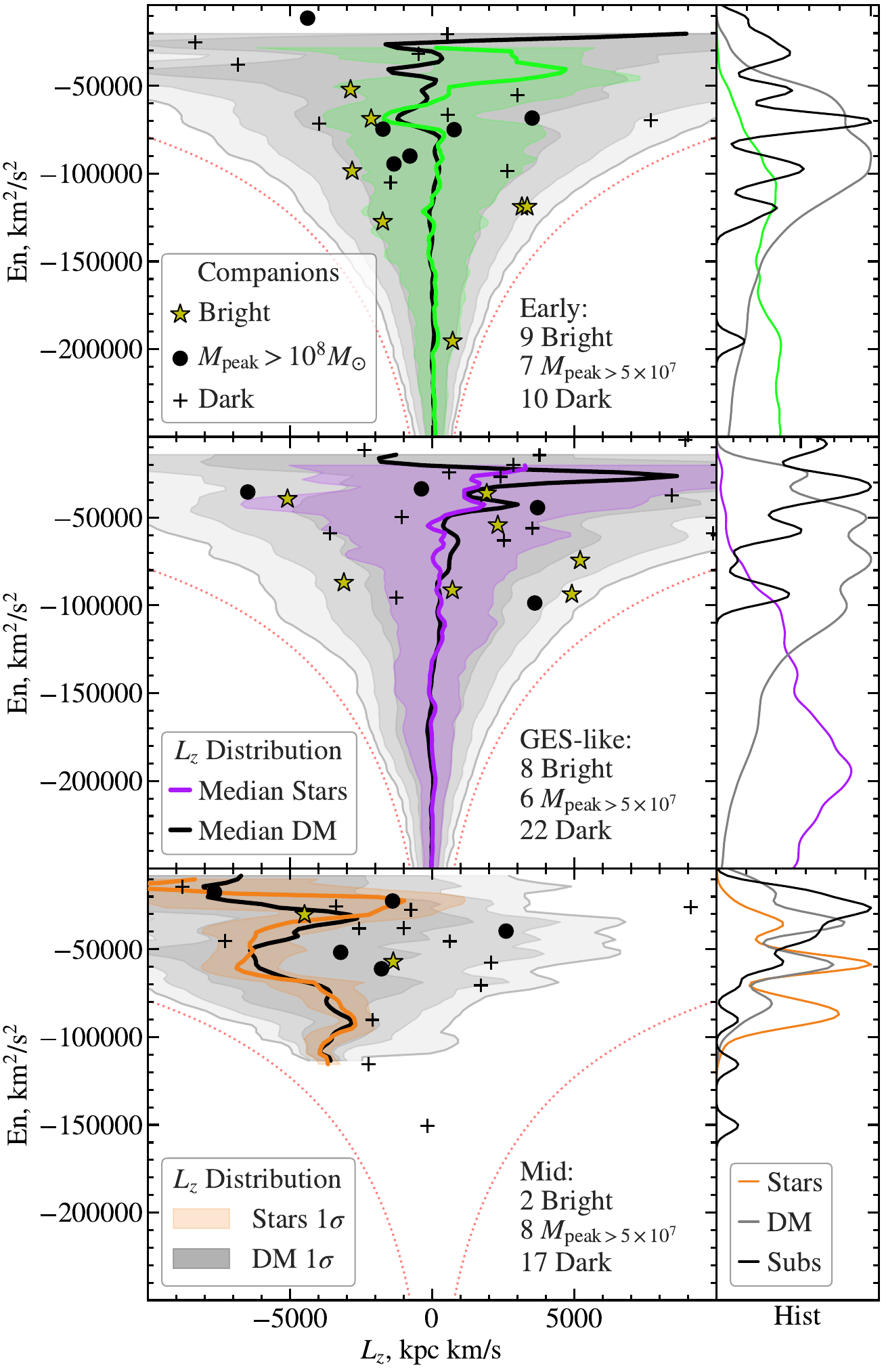}
 \vspace{-12pt}
 \caption{The position in energy and $z$-angular momentum of the primary satellite's debris,
 in grey for the dark matter and coloured for the stellar material.
 The solid lines give the median, and the $1\sigma$ (defined as $16\%-84\%$) is indicated by the shaded lines.
 \changed{With symbols, we indicate the position their companions material at the present day
 through their most bound particles at infall.
 The bright satellites are represented with star symbols,
 the dark subhaloes with peak mass greater than $10^{8}\Msol$,
 and smaller dark subhaloes with crosses (see Sec.~\ref{Sec:AuData}).}
 The top, middle, and bottom panels show examples of our Early, GES-like, and Mid analogue selections (see text for details). 
 The side histograms show the energy distribution of the primary satellite's stars (coloured), dark matter (light grey) and subhaloes (black).
\vspace{-10pt}
 }\label{fig:ELz}
\end{figure}

In this section, we study the orbital structure of the infalling groups,
\changed{considering both the present day distribution of the primary satellites stars and dark matter,
and the companions material as traced by the most bound particle at infall.
}
In Fig.~\ref{fig:ELz}, we show examples of the result of group infall at the present day in energy and $z$-angular momentum space,
which is often used to identify substructures possibly associated with debris from accretion events \citep{helmi00MappingSubstructureGalactic}.

\changed{
The accreted material of more massive accretion events typically exhibit a broader spread in IoM space 
compared to that from smaller satellites,
as can be seen in the Kraken and GES-like events in the top two panels of Fig.~\ref{fig:ELz}.
At infall, their material has a larger distribution in physical space and a higher internal velocity dispersion,
leading to a wider distribution of angular momentum.
The larger progenitors then suffer greater dynamical friction,
shedding outer layers, the majority dark matter,
at higher energies, where it shows some clumped features (as shown in grey in the side histogram).
The central material,
including the bulk of the stellar material,
sinks to lower energies, giving an extended energy distribution.
}

\changed{
Many of the companions are stripped early in the accretion alongside the outer dark matter.
Therefore,  there is limited overlap with the majority of the stellar material for the larger events.
Furthermore, there is no easily distinguishable structure in angular momentum,
with the distribution of companions consistent with that of isolated satellites at the same energy.
However, for the smaller, more recent Sagittarius-like accretion event,
the material occupies a smaller region in the $E$-$L_z$ space and shows more structure.
Their companions are not consistent with the distribution of isolated satellites,
and instead roughly follow the smaller region of the material of the primary satellite.}

\subsection{Correlation in Angular Momentum}
So far, we have focused on comparing the distribution of debris from the primary with that of the companions.
Here, we ask whether there is evidence of structure in the angular momentum space of the grouped companions.
To investigate, we consider a correlation function of the angular momentum orientation.

We calculate the angular momentum vectors $\boldsymbol{L}$ of the most bound particles 
of the subhaloes (companions) at infall.
The angular distance between two vectors is then:
\begin{equation}
  \alpha_{\boldsymbol{L}}^{i,j} = \arccos\left(\frac{\boldsymbol{L}_{i}\cdot\boldsymbol{L}_{j}}{\left|\boldsymbol{L}_{i}\right| \left|\boldsymbol{L}_{j}\right|}
  \right).
\end{equation}

Within a selected group, we find the cumulative distribution of these angular distances $H\left(\alphaL\right)$.
For an isotropic distribution:
\begin{equation}
H_{\mathrm{Isotropic}}\left(\alphaL\right)=\frac{1-\cos\left(\alphaL\right)}{2}.
\end{equation}
We define the correlation function $\omega\left(\alpha_{\boldsymbol{L}}\right)$ as:
\begin{equation}
\omega\left(\alpha_{\boldsymbol{L}}\right) = H\left(\alphaL\right)/H_{\mathrm{Isotropic}}\left(\alphaL\right).
\end{equation}
Therefore, $\omega\left(\alpha_{\boldsymbol{L}}\right)$ denotes the excess of pairs with angular separation $\alpha_{\boldsymbol{L}}$ in the data compared to an isotropic distribution.

\begin{figure}
 \includegraphics[width=\columnwidth]{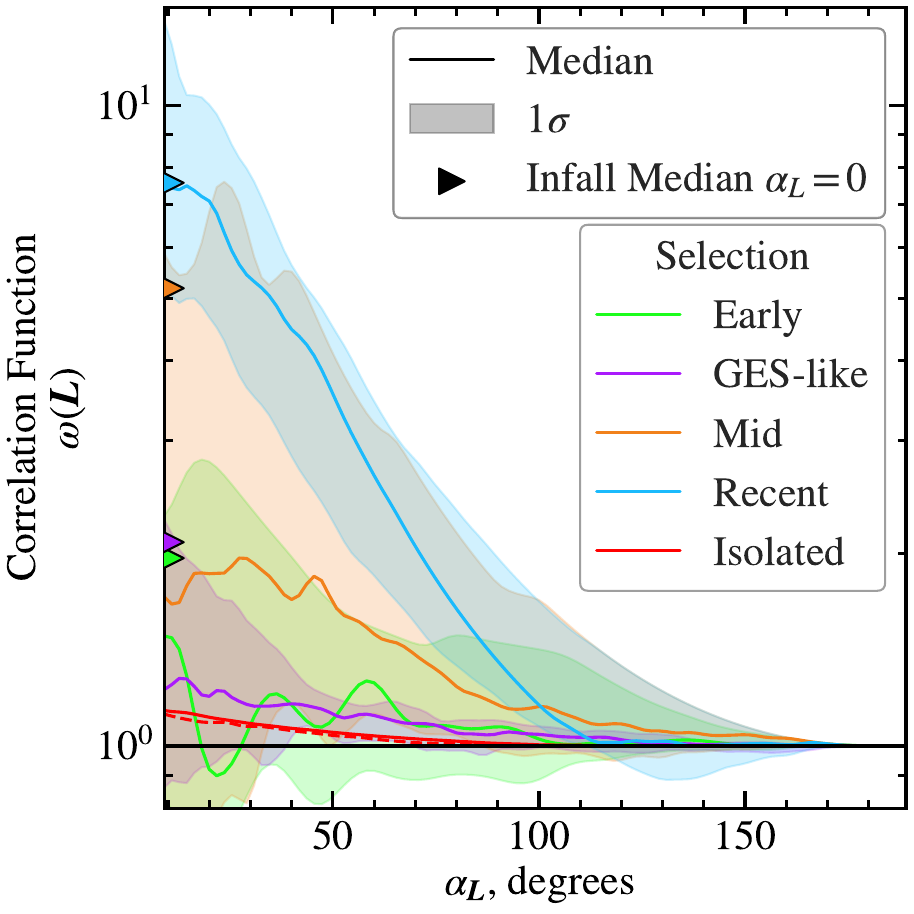}
 \vspace{-12pt}
 \caption{ 
 Correlation of selected infall groups direction of angular momentum at present day.
 The correlation function is the excess of pairs compared to an isotropic distribution on a given angular separation $\alphaL$. 
 The different colours correspond to our different selections.
 The solid line represents the median correlation of our different group samples,
 with the $1\sigma$ scatter ($16\%-84\%$ within our sample given by the shaded regions.
 \changed{The triangles at $\alpha_{L}=0^o$ correspond to the (median) correlation amplitude on the smallest scale for the different selections at the time of infall.}
 }\label{fig:L_Corr}
\end{figure}

In Fig.~\ref{fig:L_Corr}, we plot the correlation function for our selections
and isolated subhaloes (i.e. those that have not fallen in with a group).
We see that on small scales, the isolated satellites show a slight signal,
possibly indicating directions of preferential accretion, such as filaments.
The most recent selection, the LMC-like accretion events, shows a very strong signal,
implying that most satellites are orbiting on the same plane \citep[as also observed, e.g.][]{sales17IdentifyingTrueSatellites}.
The Sagittarius-like (Mid) selection also shows an excess, although of smaller amplitude 
and greater scatter, indicating the variety of events considered.
For the last major merger, there is a small correlation on small scales.  
Finally, the correlation is completely lost for the earliest accretion events, with no discernible dynamical signal remaining at the present-day.
\changed{As expected, at the time of the infall, all selections reveal a much higher degree of correlation, as indicated by the triangle symbols on the y-axis.}

The reason for the decline in amplitude \changed{of the correlation function} with time can be attributed to the change in the orbital plane, which itself evolves in time, implying that older accretion events will be more affected.
This change is exacerbated by the fact that orbits in the inner parts of a galaxy experience a potential far from spherical due to the presence of a disk. 
\changed{Furthermore, significant mergers after an accretion event can disturb its dynamics,
and reduce the correlation within the associated group \citep[e.g.][]{kanehisa23ImprintGalaxyMergers}.}

\subsection{Radial Structure}
\begin{figure}
 \includegraphics[width=\columnwidth]{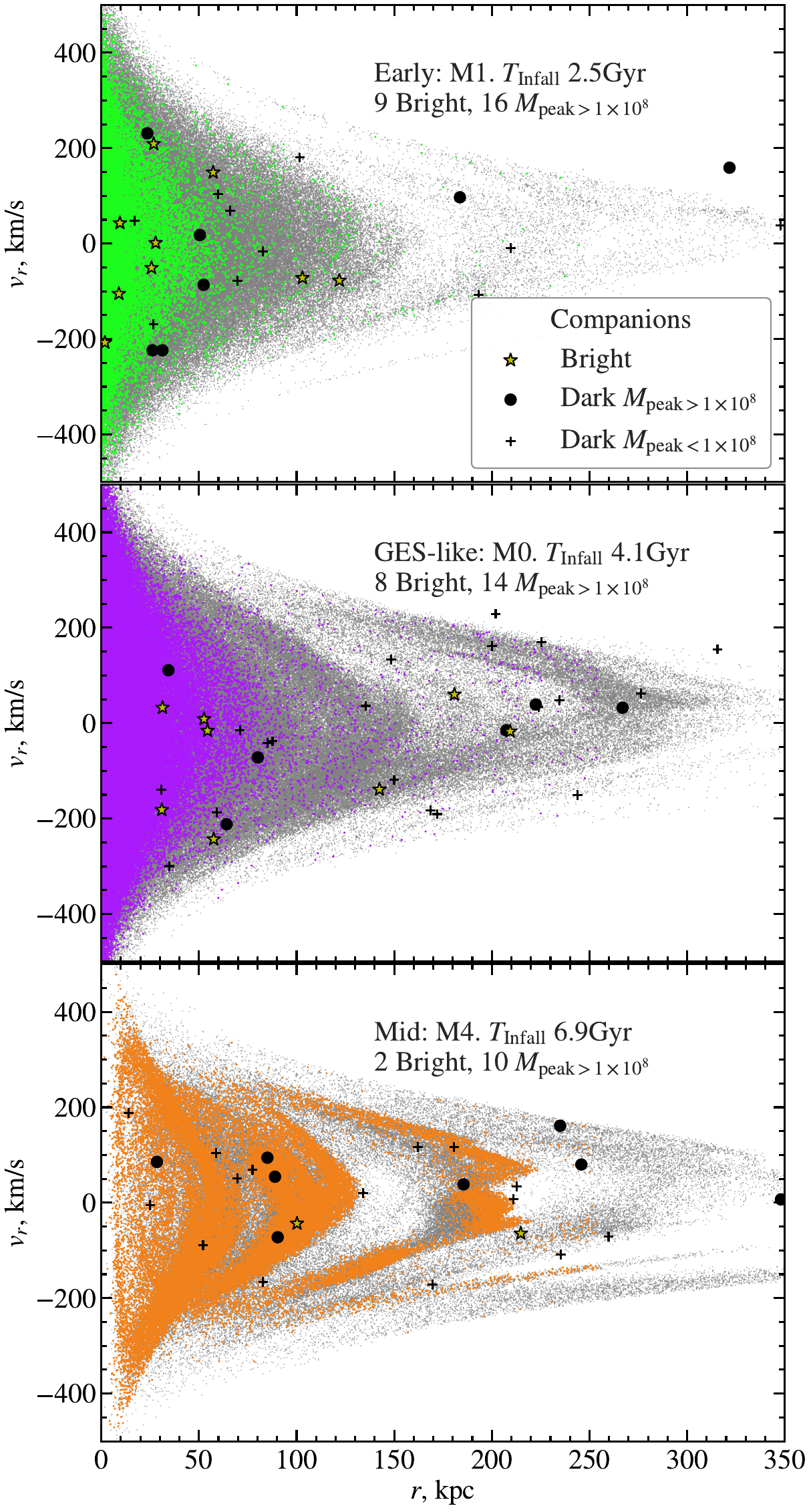}
\vspace{-12pt}
 \caption{Radial distance and radial velocity of the primary satellite's stars (coloured points) and DM (grey points).
 The top, middle, and bottom panels show examples of our Early, GES-like, and Mid analogue selections (see text for details).
 The bright satellites are yellow stars, the dark subhaloes with a peak mass greater than $10^{8}M_{\odot}$ are black circles,
 and those less than $10^{8}M_{\odot}$ are black crosses.
 }\label{fig:rvR}
\vspace{-10pt}
\end{figure}

We now explore the accreted material in radial distance and radial velocity phase space (Fig.~\ref{fig:rvR}).
Here, chevron features can be seen, relating to material lost at different epochs \citep{genina23EdgeRelationStellar},
with some correspondence with those present in the energy distributions of the material seen in  Fig.~\ref{fig:ELz}.
We see tentative evidence that some companions overlap with these chevrons,
particularly in the most distant material,
which is typically associated with the first stripping episode when the system's outskirts were lost.
This overlap is clearest in the Sagittarius-like accretion event, as seen from the bottom panel of Fig.~\ref{fig:rvR}. 

The association between the stellar material and the companions is less obvious for GES (middle panel of Fig.~\ref{fig:rvR}).
The reason for this is that a large fraction of the mass from this system was lost in one go,
preventing the view of the individual stripping events taking place at each pericentric passage in the case of a lower mass system.
The overlap is more apparent with the dark matter as this was also stripped early at higher energy,
and therefore has large apocentres and orbital periods, helping to retain greater coherence.
Similarly, there is an overlap with the stellar material from GES, which has the largest apocentres. 

The material piles up in shells at apocentre (along $v_{r}=0$). 
For the GES-like accretion event, the most distant shell and associated companions have experienced only 3 pericentres by the present time,
due to the longer orbital timescales of the higher energy orbits.
Whilst the resolution prevents a confident and detailed analysis of the debris, it is clear that the companions are largely embedded in 
the more distant material originating from GES.

\section{Discussion}\label{Sec:MW_Discuss}
In this section, we contextualise our results and discuss the possible implications for the MW.

In our simulations, we find that the Kraken-like events 
are likely accompanied by 6-10 bright companions, all destroyed in the present day.
The companion's material resides within the central $20\kpc$ of the host but is less bound than the very centrally concentrated stars of the progenitor.
The stellar material of these companions will be very old and likely to be metal-poor,
thus making them presumably difficult to distinguish from the general halo. 

We find that GES-like events are likely accompanied by a population of bright companions, with a median of around $6$.
Around half of these can be found in the solar neighbourhood, and half orbit beyond $15\kpc$ in the outer halo.
Most of these companions are presently in the higher tails of the primary satellite's energy distribution
due to the significant dynamical friction and subsequent sinking experienced by the massive progenitor.
Whilst the distribution of the primary satellite's material is too broad for clear associations,
we see evidence that the surviving companions could still be linked to the outer shells of GES,
overlapping kinematically with debris from the system's outskirts.
This possible relation indicates that should any ultra-faint dwarves overlap with the Virgo overdensity,
thought to be associated with an initial turnaround location of GES debris \citep{balbinot21LinkingNearbyStellar},
they could likely be companions of GES.

As our largest destroyed accretion event, GES probably had the largest associated companion.
\changed{In our simulated sample, $25\%$ of GES-like events have a companion larger than $1/100$  ($\sim1\times10^{7}\Msol$) in peak stellar mass.}
In our current understanding of the Milky Way assembly history,
Sequoia \citep{myeong19EvidenceTwoEarly} is the clearest candidate for having been a massive companion of GES (see Fig.~\ref{fig:MW_TMass}),
although possibly also LRL-64/Antaeus \citep{ruiz-lara22SubstructureStellarHalo,oria22AntaeusRetrogradeGroup}.
Whilst Sequoia debris was discovered near the Sun, its member stars have apocentres of $\sim 20-25 \kpc$,
which is not too different from the distance to the Virgo overdensity,
which, as stated earlier, has been associated with GES debris \citep{deason18ApocenterPileupOrigin,balbinot21LinkingNearbyStellar}.

Our Middle selection of classical satellites shows significant variety,
some of which are likely to be similar to Sagittarius.
Nearly all infall with at least one bright companion, a fraction of which still survive in the outer halo.
These companions can show a much closer relation to the primary satellite's material,
both kinematically and in some coherence with the orientation of their orbital planes.

Our Recent selection shows 5 accretion events analogous to the Magellanic group.
Although \Auriga{} predicts only a few with bright companions,
this likely results from the limitations of the simulations.
By assuming that all satellites with a mass greater than $10^{8}M_{\odot}$ are in fact bright,
we find that the median number of bright satellites increases to 7,
in line with the predictions of other works \citep{deason15SatellitesLMCmassDwarfs}.
The orbital configuration of most of these companions in the present day follows the orientation of the primary satellites in a plane,
much like what we see in the Magellanic group.

\section{Conclusion}\label{Sec:Conclusion}

Group infall has likely shaped our stellar halo.
In the \Auriga{} simulations, we have identified the likely populations of companions to key accretion events 
in our Galaxy's past: Kraken, GES, Sagittarius, and the LMC.
We find that all large accretion events are likely to be accompanied by bright companions.

The largest accretion events typically leave their companions at a greater radius and with lower binding energy.
The material of the companions of the oldest building blocks of the MW, such as Kraken, are likely to exist within the solar neighbourhood
but show no discernable coherence in their dynamic properties in the present day. 
We predict that half of the companions of GES still lie in the outer halo,
where they could be associated with the outer shells of GES.
The companions of Sagittarius can likely be associated dynamically, using, for example, phase-space diagrams such as $\left(r,v_r\right)$ or coherence in the orbital angular momenta. In all cases, we find that the companions are embedded in the material of the primary satellite, especially that deposited at large distances, but some of this material will be dark. 

The next \Gaia{} data release will significantly increase the number of stars with complete position-velocity information.
This increase could allow further smaller substructures to be discovered, perhaps even the ultra-faint companions of our most massive building blocks.
Future spectroscopic surveys, such as WEAVE and 4MOST, could also help identify these lost companions based on their distinct chemistry.
Finally, surveys such as Euclid, DESI, and LSST will shed light on the MW's outer halo,
where we predict the material of many companions to lie waiting to be discovered.

\section*{Acknowledgements}
We are grateful to Anna Genina for the useful discussions, \changed{and to the referee for a constructive report that sharpened the manuscript}. We have used simulations from the Auriga Project public data release \citep{grand24OverviewPublicData} available at {\tt https://wwwmpa.mpa-garching.mpg.de/auriga/data}.
We acknowledge financial support from a Spinoza prize to AH. 
This work used the DiRAC@Durham facility managed by the Institute for Computational Cosmology on behalf of
the STFC DiRAC HPC Facility (www.dirac.ac.uk), and part of the National e-Infrastructure.
The equipment was funded by BEIS capital funding via STFC capital grants ST/P002293/1, ST/R002371/1 and ST/S002502/1,
Durham University and STFC operations grant ST/R000832/1. 
The analysis has benefited from the use of the following packages:
Pynbody \citep{pontzen13PynbodyNBodySPH},
\textsc{AGAMA} \citep{vasiliev19AGAMAActionbasedGalaxy},
NumPy \citep{vanderwalt11NumPyArrayStructure},
matplotlib \citep{hunter07Matplotlib2DGraphics}
and jupyter notebooks \citep{kluyver16JupyterNotebooksPublishing}. 
TC thanks RC for a lifetime of support.




\bibliographystyle{mnras}
\bibliography{references.bib}


\appendix
\section{\changed{Example Analogues}}
\label{App:ExampleSelections}

\begin{figure}
 \includegraphics[width=\columnwidth]{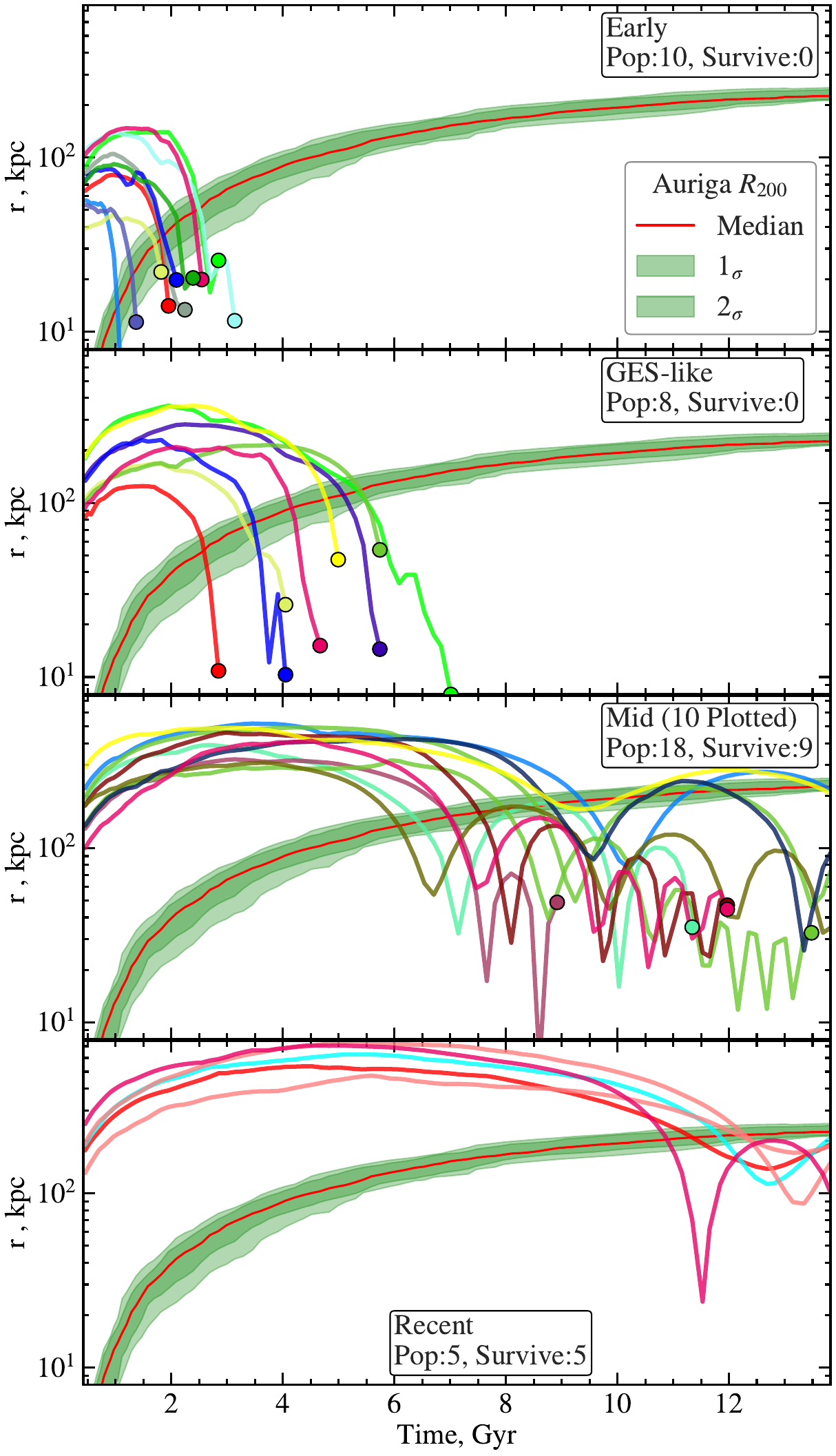}
\vspace{-12pt}
 \caption{Examples of the infall of our selection accretion events, across the Auriga halo sample.
 The different panels depict our different analogue selections,
 whose radial orbits against time are shown by the solid coloured lines.
 The point at which the subhalo was destroyed, and cannot be found in the merger trees,
 is given by the scatter points.
 The red line shows the median $R_{200}$ of our Auriga haloes, 
 with the $1\sigma$ and $2\sigma$ percentile ranges of the sample shown by green shading.
\vspace{-10pt}
 }\label{fig:AppExample}
\end{figure}

Whilst our analogue selections are relatively broad, there is general agreement amongst the samples,
as can be seen in the orbits of selected accretion events (see Fig.~\ref{fig:AppExample}).
The selections of the larger (by fractional mass) Early and GES-like accretion events
are destroyed before multiple pericenters.
The smaller Sagitarius-like ``Mid'' accretion events typically take several passages,
and some can survive on more circular outer orbits to the present day.
Finally, the Recent LMC-like accretion events have all survived a single pericenter.

\section{\changed{Resolution Effects and the Largest Companion}}
\label{app:B}

\begin{figure}
 \includegraphics[width=\columnwidth]{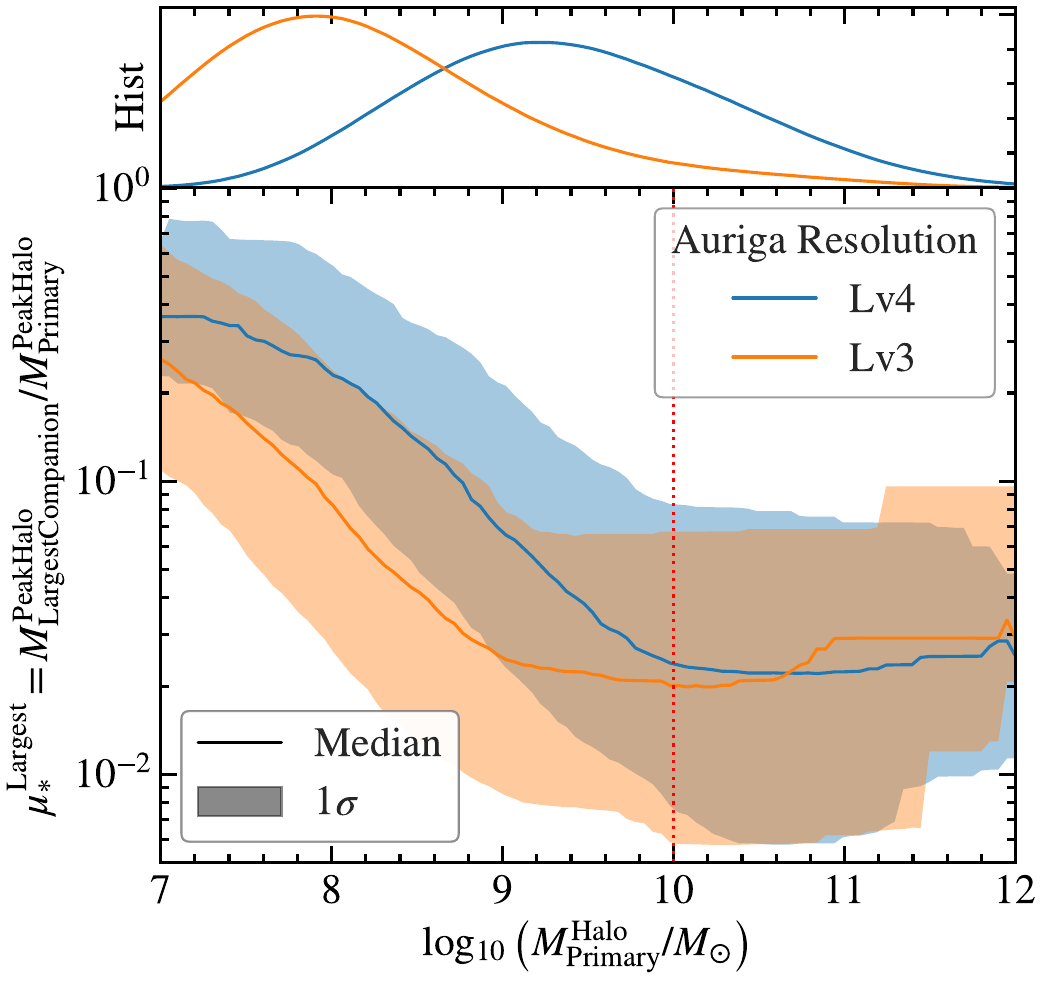}
\vspace{-12pt}
 \caption{The mass fraction of the largest companion and the primary satellite, as a function of the primary satellite's mass.
 Here we use the entire sample of group infall events for the Auriga Lv4 resolution sample (in blue),
 and the smaller Auriga Lv3 sample (in orange).
 The solid lines correspond to the median of the samples binned by primary satellite's mass,
 and the shaded regions as the $1\sigma$ ranges given by the $16\%-84\%$ percentile range.
 The additional top panel depicts the normalised histogram of the primary accretion events' mass.
\vspace{-10pt}
 }\label{fig:AppResolution}
\end{figure}

As discussed in Sec.\ref{Sec:AuData}, our simulations have finite resolution,
and in the level 4 \Auriga{} simulations this likely limits the reliable modelling of subhaloes  to $\sim4\times10^7\Msol$
( $100$ particles, see \citet{grand21DeterminingFullSatellite}).
Whilst this is more than sufficient to model the selected primary satellites,
this will affect the populations of their smaller companions,
and could conceivably affect the size of their largest companions.
To consider these effects of mass resolution,
we consider the higher resolution level 3 Auriga simulations.
Here, six of the haloes have been resimulated with a DM particle mass of $5.1\times 10^{4} M_{\odot}$
(approximately $8$ times improvement on level 4)
and an initial gas resolution element of mass $6.4\times10^{3} M_{\odot}$,
allowing subhaloes of mass greater than $\sim5\times10^6\Msol$ ($100$ particles) to be reliably resolved.

Consider the largest companions as a function of primary satellite mass in Fig.~\ref{fig:AppResolution}.
There is agreement in the two different resolution levels for primary satellites greater than $\sim10^{10}\Msol$,
indicating that the resolution has converged.
For this resolved regime, the lower end of the mass fraction distribution is  $\mu\sim5\times10^{-3}$.
At the breaking point of $10^{10}\Msol$,
this corresponds to the largest companions being of mass $\sim 4\times10^{7}\Msol$,
around the 100 particle limit for the level 4 resolution that we can expect for reliable subhalo detection.
Below this, the smallest companions can no longer be confidently resolved,
and so the distribution trends upwards.
This also happens in the level 3 resolution at $\sim2\times10^{9}\Msol$,
approximately $8\times$ lower mass, as expected.

In summary, we can consider that the largest companions of primary accretion events of greater than $10^{10}\Msol$
to be resolved in our Lv4 sample.
This includes all of our analogue selected accretion events.
Furthermore, we find that this resolution-dependent break at  $10^{10}\Msol$ has no significant dependence on the infall time of the group.

\end{document}